\documentstyle [epsf,epsfig,12pt]{article}

\newcommand{\GeV}     {\mbox{$ \,{\mathrm{GeV}}                         \, $}}
\newcommand{\mrad}    {\mbox{$ \,{\mathrm{mrad}}                        \, $}}
\newcommand{\pico}    {\mbox{$ \,{\mathrm{pb}}                        \, $}}
\newcommand{\femto}    {\mbox{$ \,{\mathrm{fb}}                        \, $}}
\newcommand{\gaga}      {$\gamma^{*}\gamma^{*}$}

\def\beqa{\begin{eqnarray}}
\def\eqa{\end{eqnarray}}
\newcommand{\kb}{\mbox{$\underline {k}$}}
\newcommand{\xb}{\mbox{$\underline {x}$}}

\newcommand{\de}{\mbox{$\frac{1}{2}$}}

\def\beq{\begin{equation}}
\def\eq{\end{equation}}

\renewenvironment{thebibliography}[1]
          {\begin{list}{[\arabic{enumiv}]}
          {\usecounter{enumiv}\setlength{\parsep}{0pt}
\setlength{\leftmargin .75cm}{\rightmargin 0pt}
        \setlength{\itemsep}{8pt} \settowidth
        {\labelwidth}{#1.}\sloppy}}{\end{list}}
\begin{document}
\input epsf

\begin{titlepage}
\vspace*{-1.5cm}

\begin{center}
\baselineskip=13pt

\vspace{2cm}

{\Large \bf  $\gamma^* \gamma^*$ total cross-section in the dipole picture
of BFKL dynamics \\}
\vskip2.5cm
{\Large Maarten Boonekamp}\\
\vskip1cm
{\it Service de Physique des Particules, DAPNIA, CEA-Saclay \\
91191 Gif sur Yvette Cedex, France}\\
\vskip1.5cm
{\Large Albert De Roeck}\\
\vskip1cm
{\it Deutsches Elektronen-Synchrotron, DESY, \\
Notkestr.85,  D-22603 Hamburg, Germany, \\
{\rm and}\\ 
CERN, CH-1211 Geneve 23, Switzerland}\\
\vskip1.5cm
{\Large Christophe Royon}\\
\vskip1cm
{\it Service de Physique des Particules, DAPNIA, CEA-Saclay \\
91191 Gif sur Yvette Cedex, France}\\
\vskip1.5cm
{\Large Samuel Wallon}\\
\vskip1cm
{\it 
Division de Physique Th\'eorique,\footnote{Unit\'e
 de Recherche des Universit\'es Paris 11 et Paris 6 Associ\'ee au CNRS} 
Institut de Physique Nucl\'eaire d'Orsay \\ 
91406 Orsay, France \\
{\rm and} \\
Laboratoire de Physique Th\'eorique des Particules El\'ementaires,  \\
Universit\'e P. \& M. Curie, 4 Place Jussieu \\
 75252 Paris Cedex 05, 
France}\\
\end{center}
\vspace*{1.5cm}
\begin{abstract} 
The total $\gamma^*\gamma^*$ cross-section is derived in the 
 Leading Order QCD dipole picture of BFKL  dynamics, and compared  
 with the one from 2-gluon exchange.
The Double Leading Logarithm approximation of the DGLAP
cross-section is found to be small in the 
phase space studied. Cross sections are calculated for 
realistic data samples at the $e^+e^-$ collider LEP and a 
future high energy linear collider.
Next to Leading order corrections to the 
BFKL evolution have been  determined phenomenologically,  
and are found to  give very large 
corrections to the BFKL cross-section, 
leading to a reduced sensitivity 
for observing BFKL. 
\end{abstract}

\end{titlepage}
\mbox{}
\setcounter{equation}{0}
\setcounter{page}{1}
\section{Introduction}
\label{introduction}
In this paper we study the 
possibility to investigate QCD pomeron effects in the high energy limit 
in virtual photon-photon scattering both at 
LEP and a future Linear Collider (LC). In the past years,
the BFKL pomeron \cite{bfkl} has been studied intensively in the 
small-$x$ regime at HERA both in the context of  proton diffractive and
fully inclusive structure functions \cite{nprw} \cite{bialas} 
and  of final state particle
flow or forward jet production \cite{albert}. The coupling
to the proton induces a non-perturbative scale in the structure function
studies and the studies of final 
states suffer from non-perturbative hadronisation 
effects. The use of a purely perturbatively calculable 
 process is much more  favourable to establish effects of BFKL
dynamics.
The cross-section of collisions of two objects with small transverse 
size is an ideal process where the BFKL approximation is 
expected to be most reliable. High energy virtual
$\gamma^* \gamma^*$ interactions at $e^+e^-$ colliders is such a process,
and has been proposed in \cite{brl,bhs,bial}
as a laboratory to 
study BFKL.

In this paper inclusive 
virtual photon scattering in $e^+e^-$ collisions at LEP 
 and a future LC  is studied. In particular the LC, with an
anticipated  luminosity 
three orders of magnitude larger than the one presently  
 at LEP, and its larger 
centre of mass (CMS) energy of  up to 1 TeV, offers an excellent 
opportunity to test BFKL dynamics.
\par
First we obtain the Leading Order (LO) BFKL cross-section using the QCD
dipole picture of BFKL dynamics. The Double Leading Logarithm
approximation of the DGLAP cross-section \cite{dglap} 
is compared
with the BFKL one. The 2-gluon approximation, where only 2 gluons are
exchanged in the $\gamma^* \gamma^*$ interactions will turn out to be 
the dominant 'background' in the region of phase space studied in 
this paper.
 We will then consider phenomenologically the effects of higher
order corrections
to the BFKL cross-section, and show that the  cross-sections
for the 2-gluon and BFKL-NLO evolutions both at the LC and LEP colliders, 
are different by a factor of two to four.

\section{$\gamma^*\gamma^*$ total cross-section in the dipole picture of BFKL
dynamics}
\label{crosssection}

\subsection{BFKL cross-section}

We analyse the $\gamma^*\gamma^*$ subprocess in the framework of the 
color dipole model \cite{mueller94, muellerpatel, muellerunitarite, nik}.
\begin{figure}
\centering
\mbox{\psfig{file=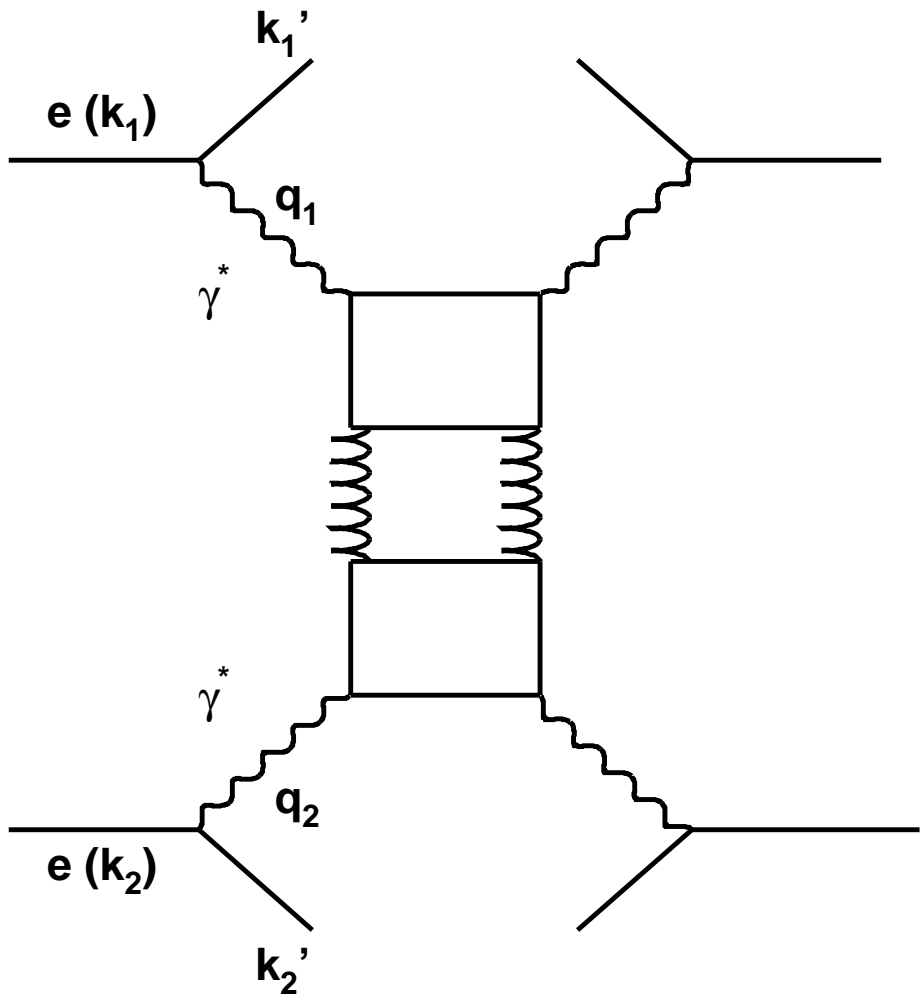,
bbllx=30pt,bblly=150pt,bburx=400pt,bbury=650pt,width=16cm,angle=0}}
\vspace{-5cm}
\caption{2-gluon exchange for the
 $\gamma^*\gamma^*$ subprocess in $e^+e^- \to e^+e^-+ X$ collisions.}
\label{grapheee}
\end{figure}
As usual, in analogy with Deep Inelastic 
Scattering (DIS) kinematics, we define the scaling variables 
which will describe the total cross-section (see Ref. \cite{brl}, and the 
scheme in Figure 1 for the definitions of the variables.)
\beq
\label{defy}
y_1 = \frac{q_1 k_2}{k_1k_2}, \quad  y_2 = \frac{q_2 k_1}{k_1k_2},
\eq
and
\beq
\label{defx}
x_1 = \frac{Q^2_1}{2q_1k_2}, \quad  x_2 = \frac{Q^2_2}{2q_2k_1},
\eq
the photons having virtualities $Q^2_1 = -q_1^2$ and $Q^2_2 = -q_2^2.$
The total energy available in the $s$ channel of the $e^+e^-$ system is
$s = (k_1 + k_2)^2.$ The energy available for the subprocess $\gamma^*\gamma^*$
is $\hat{s} = (q_1 + q_2)^2 \simeq s y_1 y_2.$
We consider the domain of large $Q^2_1,\,  Q^2_2$ (in order to be in the
 perturbative regime) and large $\hat{s}$ with the constraint
\beq
\label{ordreBFKL}
Q^2_1, \, Q^2_2 \ll \hat{s},
\eq
in order to exhibit the energy dependence of the BFKL pomeron, which will 
be described in terms of a dipole cascade. Defining 
\beq
\label{defX}
X_1 = \frac{Q^2_1}{2 q_1 q_2} \simeq \frac{Q^2_1}{\hat{s}},
 \quad X_2 = \frac{Q^2_2}{2 q_1 q_2}\simeq \frac{Q^2_2}{\hat{s}},
\eq
the total rapidity of the $\gamma^*\gamma^*$ subprocess is then given by
\beq
\label{Y}
Y = \ln \frac{1}{\sqrt{X_1 X_2}} \simeq \ln \frac{\hat{s}}{\sqrt {Q^2_1 Q^2_2}}
= \ln \frac{s y_1 y_2 }{\sqrt{Q^2_1 Q^2_2}}.
\eq
Let us first compute the $\gamma^* \gamma^*$ total cross-section in the two-gluon 
exchange approximation. Note that the QED contribution which 
corresponds to a 
quark box
coupled to four external virtual photons, is sub-leading in the high-energy 
limit, since $t$-channel exchange of two particles of spin $J$ contributes like
$s^{(2J-1)},$ and is thus dominated by gluon contributions.

In the dipole approach, one can view the two virtual photons as two dipoles,
which can scatter through the exchange of a pair of soft gluons.
This requires the knowledge of the photon wave function. The photon wave
function squared 
$\Phi_{T,L}(\underline{x},z;Q^2)$ gives the probability distribution of
finding a dipole configuration of transverse size $\underline{x},$ $z$ ($(1-z)$) being
the light-cone momentum fraction carried by the quark (antiquark). 
The subscript $T$ ($L$) corresponds to a transversally 
(longitudinally) polarized photon. Note 
that we neglect here the effect of quark masses. Heavy quark production will
be considered in a future paper.

Explicit expressions for these probability distributions can be found in 
\cite{bjorken,nik} \footnote{Note that our definition of the wave function is
such that 
$\Phi_{T,L} = \frac{2 \sqrt{2}}{N_c} \Phi^{NZ}_{T,L}$
where $\Phi^{NZ}_{T,L}$ are the corresponding wave functions squared
in reference \cite{nik}.}.
We will not use them here.

The cross-section is then obtained by convoluting 
the two probability distributions with the elementary dipole-dipole 
cross-section. The latter  corresponds to the scattering of two
color neutral objects in the eikonal approximation \cite{muellerpatel,nw}.
It reads, if one averages over the angle between the two dipoles
\footnote{Note that this cross-section is normalized so that 
$\bar{\sigma}_{DD}= \frac{N_c^2}{8}  \bar{\sigma}^{NZ}_{DD}$
where $\bar{\sigma}^{NZ}_{DD}$ is the corresponding cross-section in reference 
\cite{nik}.
},
\beq
\label{sigma_DD}
\bar{\sigma}_{DD}(\underline{x}_1,\underline{x}_2) = \alpha_s^2 \int \frac{d^2 \underline{k}}{(\underline{k}^2)^2}
\left(2-e^{\displaystyle i\underline{k}.\underline{x}_1}-e^{-\displaystyle i
\underline{k}.\underline{x}_1}\right)\left(2-e^{\displaystyle i
\underline{k}.
\underline{x}_2}-e^{-\displaystyle i\underline{k}.\underline{x}_2}
\right)
\eq
The $\gamma^* \gamma^*$ cross-section in the two-gluon 
exchange approximation then reads
\beq
\label{sigma01}
\sigma_{\gamma^*\gamma^*}(Q_1^2,Q_2^2;Y) = 
\int d^2\underline{x}_1\, dz_1 \,
\int d^2\underline{x}_2 \, dz_2 \, \Phi(\underline{x}_1,z_1;Q_1^2)
\, \Phi(\underline{x}_2,z_2;Q_2^2)\,  
\sigma_{DD}(\underline{x}_1,\underline{x}_2) \,.
\eq

Let us consider now the elementary Born cross-section $\hat{\sigma}_
{\gamma
 d}/k^2$ of the process  
\beq
\label{dgg}
d(\xb)~ g(k) \rightarrow  ~ d(\xb)
\eq
 for a dipole of transverse size $\underline{x}$
 and a soft gluon of virtuality $\underline{k}^2,$ in light-cone gauge.
In the high energy approximation, we only need its expression close to the
physical pole $k^2=0$. Thus its computation can be performed using
the equivalent photon (in this case gluon) approximation of Weizs\"acker and Williams
\cite{landau}, setting $k^2=0$ in the expression of the classical current of
a dipole \cite{nprw, these, nw}. This gives, summing over color and polarization 
of the emitted
gluon and averaging over the color of the dipole,
\beq
\label{born2}
\frac{\hat{\sigma}_{g d}}{\kb^2} =  \frac{\alpha_s N_c}{\pi} \left( 2 - 
e^{\displaystyle i\kb.\xb} - e^{
\displaystyle -i\kb.\xb} \right) \frac{1}{\kb^2}\,.
\eq
This squared quantity is related to the elementary dipole-dipole
cross-section since it can also be computed using the same 
equivalent gluon technique.

Indeed, the elementary dipole-dipole cross-section $\bar{\sigma}_{DD}$ reads
\beq
\label{sigma0}
\bar{\sigma}_{DD}(\underline{x}_1,\underline{x}_2) = \frac{(2 \pi)^2}{4 N_c^2}
\int d^2\underline{k} \frac{\hat{\sigma}_{g d}}{\kb^2}\,
\frac{\hat{\sigma}_{g d}}{\kb^2}\,,
\eq
the factor $\frac{(2 \pi)^2}{4 N_c^2}$ arising from the normalisation of the
$\underline{k}$ integration and from the definition of 
$\frac{\hat{\sigma}_{g d}}{\kb^2}$
as an averaged color quantity.

Thus, the  $\gamma^* \gamma^*$ total cross-section in the two-gluon 
exchange approximation now reads
\beq
\label{sigma02}
\sigma_{\gamma^*\gamma^*}(Q_1^2,Q_2^2;Y) = \left(\frac{\pi}{N_c^2}\right)^2
\int d^2\underline{x}_1\, dz_1 \,
\int d^2\underline{x}_2 \, dz_2 \, \Phi(\underline{x}_1,z_1;Q_1^2)
\, \Phi(\underline{x}_2,z_2;Q_2^2)\,  
\int d^2\underline{k} \frac{\hat{\sigma}_{g d}}{\kb^2}\,
\frac{\hat{\sigma}_{g d}}{\kb^2}\,.
\eq 
This will give us later the hint to exhibit the relation with the calculation
based on Feynman diagrams technique.

Following \cite{mp}, we define the Mellin-transform of the photon wave 
function as
\beq
\label{Phigamma}
\tilde{\Phi}(\gamma) = \int \frac {d^2 \underline{x}}{2 \pi} dz \, 
\Phi_{T,L}(\underline{x},z) (\underline{x}^2 Q^2)^{1 - \gamma} \,,
\eq
or equivalently
\beq
\label{inversePhigamma}
\int dz \, \Phi(\underline{x},z) = \int \frac{d \gamma}{2 i \pi} \frac{2}{\underline{x}^2} \tilde{\Phi}(\gamma) \, (\underline{x}^2 Q^2)^{-1 + \gamma}\,.
\eq
Eq. (\ref{sigma02}) then reads
\beqa
\label{sigma03}
\sigma_{\gamma^*\gamma^*}(Q_1^2,Q_2^2;Y) &=& \left(\frac{\pi}{N_c^2}\right)^2
\int d^2\underline{x}_1\, dz_1 \,
\int d^2\underline{x}_2 \, dz_2 \, \Phi(\underline{x}_1,z_1;Q_1^2)
\,\Phi(\underline{x}_2,z_2;Q_2^2)\,  
\int d^2\underline{k} \frac{\hat{\sigma}_{g d}}{\kb^2}\,
\frac{\hat{\sigma}_{g d}}{\kb^2} \nonumber \\
&=& \left(\frac{\pi}{N_c^2}\right)^2 \int d^2\underline{x}_1\, 
\int d^2\underline{x}_2\int \frac{d \gamma_1}{2 i \pi}
\int \frac{d \gamma_2}{2 i \pi} \frac{2}{\underline{x}_1^2} \tilde{\Phi}(\gamma_1)
\frac{2}{\underline{x}_2^2} \tilde{\Phi}(\gamma_2) \nonumber \\
&& \times
\int d^2\underline{k} \frac{\hat{\sigma}_{g d}}{\kb^2}\, (\underline{x}_1^2 Q_1^2)^{-1+\gamma_1} (x_2^2 Q_2^2)^{-1 + \gamma_2}
\frac{\hat{\sigma}_{g d}}{\kb^2}
\eqa
From Eq. (\ref{born2}) we have
\beq
\label{integre}
\frac{\pi}{N_c} \int \frac{d^2 \underline{x}}{\underline{x}^2} \hat{\sigma}_{g d} = \alpha_s \int \frac{d^2 \underline{x}}{\underline{x}^2} 
\left( 2 - e^{\displaystyle i\kb.\xb} - e^{
\displaystyle -i\kb.\xb} \right) = 4 \pi \alpha_s \int \frac{d x}{x}
 (1-J_0(kx))  \,.
\eq
The Mellin transform of this quantity then reads
\beq
\label{mellinsigma}
\frac{\pi}{N_c} \int \frac{d^2 \underline{x}}{\underline{x}^2} \hat{\sigma}_{g d} (x^2Q^2)^{-1 + \gamma} = 4 \pi \alpha_s v(1-\gamma) \left(\frac{k^2}{Q^2}\right)^{1 - \gamma}\eq
where (see Ref. \cite{nprw, these, nw})
\beq
\label{vgamma}
v(\gamma) = \frac{2^{-2\gamma-1}}{\gamma} \frac{\Gamma(1 - \gamma)}{\Gamma(1 + \gamma)} \,.
\eq
Thus, the  $\gamma^* \gamma^*$ total cross-section now reads
\beqa
\label{sigma04}
\sigma_{\gamma^*\gamma^*}(Q_1^2,Q_2^2;Y) &=& 4 \int \frac{d \gamma_1}{2 i \pi} 
\int \frac{d \gamma_2}{2 i \pi} \int \frac{d^2 \underline{k}}{\underline{k}^4}
4 \pi \alpha_s v(1 - \gamma_1) 4 \pi \alpha_s v(1 - \gamma_2) 
\left(\frac{k^2}{Q_1^2} \right)^{1-\gamma_1} 
\left(\frac{k^2}{Q_2^2} \right)^{1-\gamma_2} \nonumber \\
&& \hspace{-3.5cm} \times \, \tilde{\Phi}(\gamma_1)
\tilde{\Phi}(\gamma_2)
= 4\pi  \int \frac{d \gamma}{2 i \pi} 
4 \pi \alpha_s \tilde{\Phi}(\gamma) \, v(1 - \gamma) 
4 \pi \alpha_s \tilde{\Phi}(1-\gamma) \, v(\gamma) 
\left(\frac{Q_1^2}{Q_2^2} \right)^{\gamma} 
\frac{1}{Q_1^2}
\eqa
where in the last step the integration with respect to $\underline{k}$ has been performed, leading to $1-\gamma_2 = \gamma_1.$

Let us compare this calculation, based on the dipole approach (light-cone quantization), with the calculation based on Feynman diagram calculation (covariant
quantization). In the second approach one has to convolute the two off-shell
Born cross-sections $\hat{\sigma}_{\gamma g}/Q_1^2$ and 
$\hat{\sigma}_{\gamma g}/Q_2^2$
of the processes
  $\gamma(q_1)~ g(k)
 \rightarrow q ~ \bar{q}$ and $\gamma(q_2)~ g(k)
 \rightarrow q ~ \bar{q}.$ Here
the gluon is off-shell, quasi transverse, with a virtuality 
$k^2 \, \simeq \kb^2$.
The $\gamma^* \gamma^*$ total cross-section in the two-gluon 
exchange approximation then reads in this scheme
\begin{eqnarray}
\label{ktfactorization1}
Q_1^2 Q_2^2 \sigma_{\gamma^*\gamma^*}(Q_1^2,Q_2^2;Y) &=& \frac{2}{\pi^4} \int d^2 \underline{k}_1 \int d^2 \underline{k}_2  \int^1_0 \frac{dz_1}{z_1} 
\int^1_0 \frac{dz_2}{z_2} 
\hat{\sigma}_{\gamma g}\left(\frac{X_1}{z_1},\frac{\underline{k}_1^2}{Q_1^2} \right)\,
\nonumber \\
&& \hat{\sigma}_{\gamma g}\left(\frac{-X_2}{z_2},\frac{\underline{k}_2^2}{Q_2^2} \right) \, \delta^2(\underline{k}_1 - \underline{k}_2)\,,
\end{eqnarray}
where $z_1$ and $z_2$ are the light-cone momentum fractions of the exchanged 
gluon respectively
measured with respect
to $q_2^-$ and $q_1^+$
(compare with DIS where $x_{Bj}$ is the light-cone fraction with respect to 
the proton momentum).
In the light-cone frame, we get 
\beq
\label{momenta}
q_1^{\mu} = (\stackrel+{q_1^+},\stackrel-{-\frac{Q_1^2}{2 q_1^+}},
\stackrel\bot0)\, q_2^{\mu} = (\stackrel+{-\frac{Q_2^2}{2 q_2^+}},
\stackrel-{q_2^-},\stackrel\bot0)
\eq
with $q_1^+ \gg  q_1^-$ and $q_2^- \gg  q_2^+.$
This $k_T$ factorization \cite{catani,collins,levin}, allowed
by the high energy regime $Y \gg 1$ of the process, 
relies once more on the
 equivalent gluon approximation, which allowed us to compute the scattering 
in the dipole approach. 
Let us now define the double Mellin transformation in both longitudinal and transverse spaces, in order to deconvolute the integrations in Eq. (\ref{ktfactorization1}).
We set 
\beq
\label{mellinx}
\hat{\sigma}_{\omega} \left(\frac{k^2}{Q^2}\right) = \int^1_0 dx \, x^{\omega} \hat{\sigma}\left(x, \frac{\underline{k}^2}{Q^2}\right)\,,
\eq
or equivalently
\beq
\label{invmellinx}
\hat{\sigma}\left(x, \frac{\underline{k}^2}{Q^2}\right) = \int \frac{d \omega}{2 i \pi} \, x^{-\omega-1} \hat{\sigma}_{\omega} \left(\frac{k^2}{Q^2}\right)\,
\eq
for the longitudinal momenta,
and 
\beq
\label{hw}
4 \pi^2 \alpha_{e.m} \, (\sum_f e_f^2) \, h_{\omega}(\gamma) = \gamma \int^{\infty}_{0}
 \frac{d \kb^2}{\kb^2} \left(\frac{\kb^2}{Q^2}\right)^{\gamma} 
\hat{\sigma}_{\omega}\left(\frac{\kb^2}{Q^2}\right)
\eq
or equivalently  
\beq 
\label{hinv}
\hat{\sigma}_{\omega}\left(\frac{l^2}{Q^2}\right)= 4 \pi^2 \alpha_{e.m}\, 
(\sum_f e_f^2)
 \int^{ }_{ } \frac{d \gamma}{2 i \pi}
 \left(\frac{l^2}{Q^2}\right)^{-\gamma} \frac{1}{\gamma} h_{\omega}(\gamma)\,,
\eq
for the transverse degrees of freedom.
The index $f$ runs over the light quark flavors, $u,d,s$. The flavor $c$ will
be considered in an incoming paper. 
Thus, we take $\sum_f e_f^2 = 2/3$ for the 
numerical studies in this paper.
Eq. (\ref{ktfactorization1}) now reads
\beqa
\label{ktfactorization2}
Q_1^2 Q_2^2 \sigma_{\gamma^*\gamma^*}(Q_1^2,Q_2^2;Y) &=& \frac{2}{\pi^4} 
\int d^2 \underline{k}_1 
\int d^2 \underline{k}_2   \, (4 \pi^2 \alpha_{e.m} \, \sum_f e_f^2)^2 
\int \frac{d \gamma_1}{2i\pi} 
\left(\frac{\underline{k}_1^2}{Q_1^2}\right)^{-\gamma_1} 
\frac{h_0(\gamma_1)}{\gamma_1} \nonumber \\
&& \int \frac{d \gamma_2}{2i\pi} 
\left(\frac{\underline{k}_2^2}{Q_2^2}\right)^{-\gamma_2} 
\frac{h_0(\gamma_2)}{\gamma_2} \,
\delta^2(\underline{k}_1 - \underline{k}_2) \nonumber \\
&=& \frac{2\pi}{\pi^4} \, (4 \pi^2 \alpha_{e.m} \, \sum_f e_f^2)^2  Q_2^2
\int \frac{d \gamma}{2i\pi}  \frac{h_0(\gamma)}{\gamma}
\, \frac{h_0 (1-\gamma)}{1-\gamma} \left(\frac{Q_1^2}{Q_2^2}\right)^{\gamma}\,.
\eqa
$h_{T,L} \equiv h_{(\omega=0)T,L}$ were computed in Ref. \cite{catani} and are given by
\beq
\label{defh}
\left(\begin{array}{c}
h_T \\ h_L \end{array} \right) = \frac{\alpha_s}{ 3 \pi \gamma} 
\frac{(\Gamma(1 - \gamma) \Gamma(1 + \gamma))^3}{\Gamma(2 - 2\gamma) \Gamma(2 +
 2\gamma)} \frac{1}{1 - \frac{2}{3} \gamma} \left( \begin{array}{c} (1 +
 \gamma)
(1 - \frac{\gamma}{2}) \\ \gamma(1 - \gamma) \end{array} \right).
\eq
Now one can check that both formalisms give the same result: in Ref. \cite{mp} it was checked that 
\beq
\label{equivmp}
4 \pi \alpha_s \, \tilde{\Phi}(\gamma) v(1-\gamma) = 2 \sqrt{2} \alpha_{e.m} \frac{h(\gamma)}{\gamma} 
\sum_f e_f^2\,,
\eq
taking into account the difference of normalisation (see the footnotes at the
beginning of this part). 
This is exactly what we need, since Eqs. 
(\ref{sigma04}) and (\ref{ktfactorization2})
are identical when taking into account Eq. (\ref{equivmp}).
The result (\ref{equivmp}) only means that extracting a soft gluon from the
virtual photon can be equivalently computed by convoluting
 the distribution of dipoles
in the photon  with the  elementary gluon dipole cross-section or by
 calculating the Feynman
diagram describing the off-shell $\gamma(q_1)~ g(k)
 \rightarrow q ~ \bar{q}$ born cross-section.

Let us now consider the computation of the $\gamma^* \gamma^*$ total 
cross-section when resumming the $(\alpha_s \log s)^{n>0}$ contributions. 
In the dipole 
approach these terms appear when the relative rapidity of the two photons 
is large enough so that both photons can be considered to be made of
dipoles. At leading order, in the center of mass frame,
the two excited dipoles extracted from 
the two photons scatter through the exchange of a pair of soft gluons.
But due to the frame invariance of the process, which is closely related to
the conformal invariance of the dipole cascade \cite{nw}, the process
can also be viewed differently. Consider the frame where the right moving 
photon 1
has almost all the available rapidity while the left-moving photon has 
only enough rapidity to make it move relativistically (but not enough 
to add gluons to its wavefunction) \cite{muellersalam,kmw,nw}.
In this frame, the photon 2, which makes the original dipole 2,
scatters an excited dipole extracted from the fast right moving photon 1.
 Following Ref. \cite{mueller94, muellerpatel},
we define 
$n(\underline{x},\underline{x}',\tilde{Y})$ such that 
\beq
\label{defn}
N(\underline{x}',Y) = \int d^2 \underline{x} \int^1_0 dz_1
\,  \Phi(\underline{x},z_1) \, n(\underline{x},\underline{x}',\tilde{Y})
\eq
is the density of dipoles
 of transverse size $\underline{x}'$, where
 the momentum fraction of the softest of the two gluons (or quark or 
antiquark) which compose the
 dipole
 is larger or equal to $e^{-Y}.$ $\tilde{Y}$ is the relative rapidity with
respect to the heavy quark given by $\tilde{Y} = Y + \ln z_1.$
The leading order $\gamma^* \gamma^*$ total 
cross-section then reads
\begin{eqnarray}
\label{sigmaL01}
&~& \sigma_{\gamma^*\gamma^*}(Q_1^2,Q_2^2;Y) = 
\int d^2\underline{x}_1\, dz_1 \,
\int d^2\underline{x}_2 \, dz_2 \, \Phi(\underline{x}_1,z_1;Q_1^2)
\, \Phi(\underline{x}_2,z_2;Q_2^2)\, \nonumber \\ 
&& n(\underline{x}_1,\underline{x}'_1,\tilde{Y}_1) \, 
n(\underline{x}_2,\underline{x}'_2,\tilde{Y}_2)
\sigma_{DD}(\underline{x}_1,\underline{x}_2) \,,
\end{eqnarray} 
where $\sigma_{DD}(\underline{x}_1,\underline{x}_2)$ is the unaveraged (with respect to orientation) elementary dipole-dipole cross-section (see Eq. (A.33) of
Ref. \cite{nw}), which has a non trivial angular dependence to be taken into
account if one is interested in azimutal distributions.

The rapidities $\tilde{Y}_1$ and $\tilde{Y}_2$ are such that 
$\tilde{Y}=\tilde{Y}_1 +
 \tilde{Y}_2.$
In order to get the expression for $n(\underline{x},\underline{x}',
\tilde{Y})$, one relies on the
global conformal invariance of the dipole emission kernel, related to the
 absence of
 scale. This distribution can be expanded on the basis of conformaly
 invariant
 three points
 holomorphic and antiholomorphic correlation functions 
\cite{lipatov86, polyakov}.
 Introducing complex 
coordinates in the two-dimensional transverse space
\beqa
&&\underline{\rho}= (\rho_x,\rho_y)\\
&&\rho = \rho_x + i\rho_y \mbox{ and } \rho^* = \rho_x - i\rho_y,
\eqa
the complete set of eigenfunctions $E^{n,\nu}$ of the dipole emission kernel 
is
\beq
\label{defE}
E^{n, \nu}(\underline{\rho}_{10},\underline{\rho}_{20}) = (-1)^n\left(\frac{
\rho_{12}}{\rho_{10} \rho_{20}} \right)^h \left(\frac{\rho^*_{12}}{\rho^*_{10}
 \rho^*_{20}} \right)^{\bar h},
\eq
with the conformal weights
\beqa
h &=& \frac{1-n}{2} + i \nu \nonumber \\
\bar h &=& \frac{1+n}{2} + i \nu \,,
\eqa
where $n$ is integer and $\nu$ real.
This set constitutes a unitary irreducible representation of  SL(2,{\bf C})
 \cite{gelfand}.

The Mellin transform of $n(\underline{x},\underline{x}',\tilde{Y})$
with respect to $\tilde{Y}$ is defined by
\beq
\label{mellinomega}
n(\underline{x},\underline{x}',\tilde{Y}) = \int \frac{d \omega}{2 i \pi} e^{\omega \tilde{Y}} n_{\omega}(\underline{x},\underline{x}').
\eq
In this double Mellin space (one for longitudinal degrees of freedom, one for transverse), the dipole distribution is diagonal. One gets (see Eq. (2.65) in Ref.\cite{nw})
\beq
\label{ntotal1}
n(\underline{x},\underline{x}',\tilde{Y}) = \sum_{n=-\infty}^{+\infty} \int^{+\infty}_{-\infty} \frac{d \nu}{2 \pi} \frac{|x|}{|x'|} \left(\frac{x^* x'}{x {x'}^*}
\right)^{n/2} \left|\frac{x'}{x}\right|^{-2 i \nu} \exp\left(\frac{2 \alpha
 N_c}
{\pi} \chi(n,\nu) \tilde{Y}\right)\,,
\eq
where
\beq
\label{chinnu}
\chi(n,\nu) = \psi(1) - \de \psi\left(\frac{|n|+1}{2} + i \nu\right) -\de \psi
\left(\frac{|n|+1}{2} - i \nu\right) = \psi(1) - {\rm Re} \, 
\psi\left(\frac{|n|+1
}{2} + i \nu\right).
\eq
The
$\gamma^* \gamma^*$ total 
cross-section then reads
\beqa
\label{sigmaLO2}
\hspace{-1cm} \sigma_{\gamma^*\gamma^*}(Q_1^2,Q_2^2;Y) &=& 
\int d^2\underline{x}_1\, dz_1 \,
\int d^2\underline{x}_2 \, dz_2 \, \Phi(\underline{x}_1,z_1;Q_1^2)
\, \Phi(\underline{x}_2,z_2;Q_2^2)\,  \nonumber \\
&&  \hspace{-5cm} \times \,\frac{\pi \,
 \alpha_s^2}{2} \,  \sum_{n=-\infty}^{+\infty}
 \int_{-\infty}^{+\infty} \frac{d \nu}{2 \pi} \frac{1 + (-1)^n}{\left( \nu^2 + \left(\frac{n-1}{2}
 \right)^2
 \right) \left( \nu^2 + \left ( \frac{n+1}{2} \right)^2 \right)}   \left(
\frac{x_1^* x_2}{x_1 {x_2}^*}\right)^{n/2} |x_1|^{1 +2 i \nu} \,
|x_2|^{1 -2 i \nu} \,.
\eqa
The elementary dipole-dipole cross section $\sigma_{DD}$ can be expressed
as (see \cite{nw})
\beq
\label{sigmaangle}
\sigma_{DD}(x_1,x_2)=  4 \alpha_s^2 \frac{|x_1| |x_2|}{16} 
\sum_{n=-\infty}^{+\infty}
 \int_{-\infty}^{+\infty} d \nu \,  \frac{1 + (-1)^n}{\left( \nu^2 + \left(
\frac{n-1}{2} \right)^2 \right) \left( \nu^2 + \left ( \frac{n+1}{2} \right)^2
 \right)}
\left(\frac{x_1 x_2^{*}}{x_1^* x_2}\right)^{n/2} \left|\frac{x_1}{x_2}
\right|^{-2 i
 \nu}.
\eq
Defining
 \beq
\label{defnquant}
n(\underline{x}_1,\underline{x}_2)
= \sum_{n=-\infty}^{+\infty} \int_{-\infty}^{+\infty} \frac{d \nu}{2 \pi}
 \, n_{\{n,\nu\}}(\underline{x}_1,\underline{x}_2) \,
 \exp \left(\frac{2 \alpha_s N_c}{\pi} \chi(n,\nu) \tilde{Y} \right),
\eq
and 
 \beq
\label{sigmanquant}
\sigma_{DD}(\underline{x}_1,\underline{x}_2)
= \sum_{n=-\infty}^{+\infty} \int_{-\infty}^{+\infty} \frac{d \nu}{2 \pi}
 \, \sigma_{DD\{n,\nu\}}(\underline{x}_1,\underline{x}_2) \,
\eq
the equation (\ref{sigmaLO2}) can be written as
\beqa
\label{sigmaLO3}
&&\hspace{-.5cm}  \sigma_{\gamma^*\gamma^*}(Q_1^2,Q_2^2;Y) \nonumber \\ 
&& \hspace{-.4cm}= \int \! d^2\underline{x}_1\, dz_1 
\int \! d^2\underline{x}_2 \, dz_2 \, \Phi(\underline{x}_1,z_1;Q_1^2)
\, \Phi(\underline{x}_2,z_2;Q_2^2)\, \sum_{n=-\infty}^{+\infty}
 \int_{-\infty}^{+\infty} \frac{d \nu}{2 \pi}
\sigma_{DD\{n,\nu\}}(\underline{x}_1,\underline{x}_2) \, n_{\{n,\nu\}}(\underline{x}_1,\underline{x}_1) \nonumber \\
&&  \hspace{-.4cm}= \int \! d^2\underline{x}_1\, dz_1 
\int \! d^2\underline{x}_2 \, dz_2 \, \Phi(\underline{x}_1,z_1;Q_1^2)
\, \Phi(\underline{x}_2,z_2;Q_2^2)\, \sum_{n=-\infty}^{+\infty}
 \int_{-\infty}^{+\infty} \frac{d \nu}{2 \pi}
\sigma_{DD\{n,\nu\}}(\underline{x}_1,\underline{x}_2) \nonumber \\
&& \times \, \exp \left(\frac{2 \alpha_s N_c}{\pi} \chi(n,\nu) \tilde{Y} \right) \! .
\eqa
In the case where angular-averaged cross-section are considered, 
one can make $n=0$
in the previous expression. 
Setting $\gamma = \frac{1}{2} + i \nu$ in order
to write down expressions in term of the anomalous dimension, one gets for the 
averaged elementary cross-section
\beq
\label{resultatequiv}
\bar{\sigma}_{DD}(\underline{x}_1,\underline{x}_2) = \int^{+\infty}_{-\infty} 
\frac{d \nu}{2 \pi}  \sigma_{DD\{0,\nu\}}(\underline{x}_1,\underline{x}_2) = 
\frac{\alpha_s^2}{2} \int^{\frac{1}{2}+i\infty}_{\frac{1}{2} -i\infty} 
d \gamma \, (x_1^2)^{\gamma} ({x_2}^2)^{1-\gamma} \frac{1}{\gamma^2(1-\gamma)^2},
\eq
and thus
\beqa
\label{sigmaLO4}
 && \sigma_{\gamma^*\gamma^*}(Q_1^2,Q_2^2;Y)
= \int \! d^2\underline{x}_1\, dz_1 
\int \! d^2\underline{x}_2 \, dz_2 \, \Phi(\underline{x}_1,z_1;Q_1^2)
\, \Phi(\underline{x}_2,z_2;Q_2^2)\,  \frac{d \gamma}{2 \pi} 
(x_1^2)^{\gamma} ({x_2}^2)^{1-\gamma} 
\frac{\pi \alpha_s^2}{\gamma^2(1-\gamma)^2} \nonumber \\
&& \hspace{1cm} \times \, \exp \left(\frac{2 \alpha_s N_c}{\pi} \chi(\gamma) 
\tilde{Y} \right) \nonumber \\
&&= \int \frac{d \gamma}{2 \pi}  \tilde{\Phi}(\gamma)
\tilde{\Phi}(1-\gamma) \frac{4 \pi^3 \alpha_s^2}{\gamma^2(1-\gamma)^2}
\frac{1}{Q_2^2} \left(\frac{Q_2^2}{Q_1^2}\right)^{\gamma}\,
\exp \left(\frac{2 \alpha_s N_c}{\pi} \chi(\gamma) \tilde{Y} \right)
\eqa
with $\chi (\gamma) = \chi (0, i(\frac{1}{2} - \gamma))$.
Using Eqs. (\ref{equivmp}) and (\ref{vgamma}) one immediately obtains
 \beqa
\label{sigmaLO5}
&&\hspace{-.5cm}Q_1^2 Q_2^2 \sigma_{\gamma^*\gamma^*}
 (Q_1^2,Q_2^2;y_1,y_2) \nonumber \\
&& = 
 32 \alpha_{e.m}^2 \pi \, 
(\sum_f e_f^2)^2
\int \frac{d \gamma}{2 i \pi}  \, \frac{h_{\omega_p}(\gamma)}{\gamma}
\,  \frac{h_{\omega_p}(1-\gamma)}{1 - \gamma} Q_2^2
 \left(\frac{Q_1^2}{Q_2^2}\right)^{\gamma} 
\exp{\frac{2\alpha_s N_c}{\pi} \chi(\gamma)  Y}.\nonumber \\
\eqa
We have made the approximation $Y = \tilde{Y},$ neglecting the rapidity
taken by the quarks (see Ref. \cite{bialasb} for an interesting discussion of 
this effect).
Note that the integrand is symetrical with respect to $\gamma \to 1 -\gamma.$
Moreover one can easily check that $\frac{h(\gamma)}{\gamma} =
\frac{h(1-\gamma)}{1-\gamma}$ for each of the two 
polarizations.

Defining the flux factors
\begin{eqnarray}
&~& t_1 = \frac{1 + (1-y_1)^2}{2}, \quad l_1=1-y_1 
\\ &~& t_2 = \frac{1 + (1-y_2)^2}{2}, \quad l_2=1-y_2 \,,
\end{eqnarray}
the contribution of this $\gamma^*\gamma^*$ subprocess to the $e^+e^-$
total cross-section is
\beqa
\label{eeBFKL}
&&\hspace{-1.5cm}d\sigma_{e^+e^-}
 (Q_1^2,Q_2^2;y_1,y_2) = \left(\frac{\alpha_{e.m}}{\pi}\right)^2  \, 
(\sum_f e_f^2)^2
\frac{d Q_1^2}{Q_1^2} \frac{d Q_2^2}{Q_2^2} \frac{d y_1}{y_1} \frac{d y_2}{y_2}
\sigma_{\gamma^*\gamma^*}
 (Q_1^2,Q_2^2;Y) \nonumber\\
&& \hspace{-.5cm}= \frac{4}{9}
\frac{32 \alpha_{e.m}^4}{\pi}\,\frac{d Q_1^2}{Q_1^2} \frac{d Q_2^2}{Q_2^2} 
\frac{d y_1}{y_1} \frac{d y_2}{y_2}
\int \frac{d \gamma}{2 i \pi} 
 \left[l_1 \frac{h_L(\gamma)}{\gamma} +
 t_1 \frac{h_T(\gamma)}{\gamma}\right] 
\nonumber \\
&&\hspace{-.5cm} \times \, \left[l_2 \frac{h_L(1-\gamma)}
{1-\gamma} +
t_2 \frac{h_T(1-\gamma)}{1-\gamma}\right] \, 
\frac{1}{Q_1^2}
\, \left(\frac{Q_1^2}{Q_2^2}\right)^{\gamma} 
\exp{\frac{2\alpha_s N_c}{\pi} \chi(\gamma)  Y}\,
\nonumber \\
\eqa
where $Y$, $y_1$, and $y_2$ are related by formula \ref{Y}.
This result agrees with that of Ref. \cite{brl} and \cite{bhs}.
In the kinematical domain where $Q_1$ and $Q_2$ are of the same order,
the $\gamma$ integration can be performed by a saddle point approximation,
the saddle being located very close to $1/2$ (on the left if $Q_1>Q_2,$ on the
right if $Q_1<Q_2$).
Finally we get
\beqa
\label{eeBFKLa}
&&\hspace{-1.5cm}d\sigma_{e^+e^-}
 (Q_1^2,Q_2^2;y_1,y_2) = \frac{4}{9} \left(\frac{\alpha_{e.m}^2}{16}\right)^2  \, 
 \alpha_s^2 \, \pi^2 \sqrt{\pi} \,
\frac{d Q_1^2}{Q_1^2} \frac{d Q_2^2}{Q_2^2} \frac{d y_1}{y_1} 
\frac{d y_2}{y_2}
\frac{1}{Q_1 \, Q_2}
\, 
\frac{e^{\displaystyle \frac{4\alpha_s N_c}{\pi} Y \ln 2 }}{\sqrt{\frac{14
\alpha_s N_c}{\pi} Y \zeta(3)}} \, \nonumber \\
&& \times \, e^{\displaystyle -\frac{\ln^2 \frac{Q_1^2}{Q_2^2}}{\frac{56 
\alpha_s N_c}{\pi} Y \zeta(3)}}
 \left[2 l_1  +
9 t_1 \right] \, \left[2 l_2  +
9 t_2 \right] \,,
\eqa
where we have neglected the dependence of $h_{T,L}(\gamma_s)/\gamma_s$
with respect to $Q_1/Q_2,$ setting $\gamma=1/2.$ This 
formula agrees with the one calculated in~\cite{brl} 
and  will be used 
 in
the following to obtain the BFKL cross-sections after integration over the
kinematical variables.

\subsection{Double Leading Log and 2-gluon cross-sections}
Let us compare this cross-section with the cross-section obtained in the Double
Leading Log
approximation of the DGLAP cross-section, 
valid for $Q_1/Q_2$ far from 1.
It corresponds to replacing $\chi(\gamma)$ and $h_{T,L}$ by their
 dominant singularity
at $\gamma=0,$ corresponding to the collinear singularity respectively
of the BFKL kernel and  of impact factors, the last one reducing
then to the usual coefficient functions of 
the Operator Product Expansion.
The dominant singularities of $h_{T,L}$ when $\gamma \to 0$ are given by
\beq
\label{h0}
\left(\begin{array}{c}
h_T \\ h_L \end{array} \right) = \frac{\alpha_s}{ 3 \pi \gamma} 
\left(\begin{array}{c} 1 \\ \gamma \end{array} \right).
\eq
In the double logarithmic approximation, one sums up terms of the type
$\sum_{p \geq 0} (\alpha_s \, \ln Q_1^2/Q_2^2 \, Y)^p,$ neglecting terms with higher
powers in $\alpha_s$ of the type 
$\sum_{p \geq n \geq 0} \alpha_s^p \, \ln^{(p-1)} Q_1^2/Q_2^2 \, Y^n$ which 
would correspond to Next to Leading Order in $Q^2.$
Thus, we only keep here the contribution corresponding to the exchange of two 
 transversally polarized photons, since the longitudinal contribution (as well
as the constant term in the expansion of $h_T$ (see Eq. (\ref{devh})) is
less singular in $\gamma$ space, which leads to a decrease of the power in
$\ln   Q_1^2/Q_2^2$ (by 1 for one longitudinal photon, by 2 for two 
longitudinal photons). Taking into account these terms could be done 
consistently when including $NLQ^2$ (if one includes one longitudinal photon)
and $NNLQ^2$ (if one includes two longitudinal photons). This will be discussed
in an incoming paper.

Thus, this yields
\beqa
\label{eeDGLAPa}
&&d\sigma^{DGLAP-DLL}_{e^+e^-}
 (Q_1^2,Q_2^2;y_1,y_2) = \frac{4}{9}\left(\frac{\alpha_{e.m}}{\pi}\right)^2 
\frac{d Q_1^2}{Q_1^2} \frac{d Q_2^2}{Q_2^2} \frac{d y_1}{y_1} \frac{d y_2}{y_2}
\sigma_{\gamma^*\gamma^*}
 (Q_1^2,Q_2^2;Y) \nonumber\\
&& \hspace{-.5cm}=
\frac{128 \alpha_{e.m}^4}{9\pi}\,\frac{d Q_1^2}{Q_1^2} \frac{d Q_2^2}{Q_2^2} 
\frac{d y_1}{y_1} \frac{d y_2}{y_2}
\int \frac{d \gamma}{2 i \pi} 
 t_1 t_2 
\frac{\alpha_s^2}{9 \pi^2 \gamma^4}
\frac{1}{Q_1^2}
\, \left(\frac{Q_1^2}{Q_2^2}\right)^{\gamma} 
\exp{\frac{\alpha_s N_c}{\pi \gamma}  Y},
\nonumber \\
\eqa
A saddle point approximation gives for the $\gamma$ integration
\beqa
\label{eeDGLAP}
&&\hspace{-1.5cm}d\sigma^{DGLAP-DLL\,asymp}_{e^+e^-}
 (Q_1^2,Q_2^2;y_1,y_2) = \left(\frac{8 \, \alpha_{e.m}^2\alpha_s}{9 \pi^2 }\right)^2 \,  \sqrt{\pi} \,
\frac{d Q_1^2}{Q_1^2} \frac{d Q_2^2}{Q_2^2} \frac{d y_1}{y_1} \frac{d y_2}{y_2}
\frac{1}{Q_1^2} \, \,  \nonumber \\
&& e^{2 \sqrt{\frac{\alpha_s N_c}{\pi} Y \ln \frac{Q_1^2}{Q_2^2}}}
\, \frac{\left(\ln \frac{Q_1^2}{Q_2^2} \right)^{5/4}}{\left(\frac{\alpha_s N_c}{\pi} Y \right)^{7/4}}
 \times  
 t_1
 t_2
 \,,
\eqa
 the 
saddle-point being located at $\gamma_s = \sqrt{\frac{\alpha_s N_c}{\pi Y \ln 
\frac{Q_1^2}{Q_2^2}}}.$ This asymptotic formula requires $\gamma_s$ to be very 
small. In fact this region is far from being reached experimentally, and one 
faces the same problem as in DIS (see Ref. \cite{these,nprsw} for discussion of the DIS case).
 In the experimental regime which can be reached by LEP and LC, the correct 
way is to write down an expansion of the exponential part 
of formula (\ref{eeDGLAPa}) in terms of Bessel function.
\begin{eqnarray}
\exp \left[ \gamma \ln \left( \frac{Q_1^2}{Q_2^2} \right) +
\frac{\alpha_S N_C}{\pi \gamma}Y \right]
= J_0 (z) + \sum_{k=1}^{\infty} \left[ \left( \frac{c}{\gamma}
\right)^k + \left( \frac{-c}{\gamma} \right)^{-k} \right] J_k(z)
\end{eqnarray}
where
\begin{eqnarray}
&~& c = i \sqrt{\frac{\alpha_S N_c Y}{\pi \ln(Q_1^2/Q_2^2)}}\\
&~& z= -2i \sqrt{\frac{\alpha_S N_C}{\pi} Y \ln(Q_1^2/Q_2^2)} \,.
\end{eqnarray}
This allows to separate the integral in $\gamma$ from the integrals over
the kinematical variables. Formula (\ref{eeDGLAPa}) then reads
\beqa
\label{eeDGLAPb1}
&& d\sigma_{e^+e^-}
 (Q_1^2,Q_2^2;y_1,y_2) = 
\frac{128 \alpha_{e.m}^4 \alpha_s^2}{81\pi^3}\,\frac{d Q_1^2}{Q_1^2} \frac{d Q_2^2}{Q_2^2} 
\frac{d y_1}{y_1} \frac{d y_2}{y_2}
\int \frac{d \gamma}{2 i \pi} 
 t_1 t_2 \frac{1}{\gamma^4}
\frac{1}{Q_1^2} \nonumber \\
&& \times
\left[ J_0 (z) + \Sigma_{k=1}^{\infty} \left( \left( \frac{c}{\gamma}
\right)^k + \left( \frac{-c}{\gamma} \right)^{-k} \right) J_k(z) \right].
\nonumber \\
\eqa
Closing the contour of the $\gamma$ integration to the left and neglecting
 higher twist contribution arising from the remaining integration from
$\gamma_0 - i \infty$ to $\gamma_0 + i \infty$ (with $\gamma_0 <0$),
and using the Cauchy theorem, one gets
\beqa
\label{eeDGLAPb2}
&& \hspace{-.5cm} d\sigma_{e^+e^-}
 (Q_1^2,Q_2^2;y_1,y_2) = 
\frac{128 \alpha_{e.m}^4 \alpha_s^2}{81\pi^3}\,\frac{d Q_1^2}{Q_1^2} \frac{d Q_2^2}{Q_2^2} 
\frac{d y_1}{y_1} \frac{d y_2}{y_2} t_1 t_2
\frac{1}{Q_1^2} 
\, J_3 (z) (-c)^{-3} \nonumber \\
&&  \hspace{-.5cm} = (\frac{16 \alpha_{e.m}^2 \alpha_s}{9 \pi})^2   \frac{1}{\pi Q_1^2}
\,\frac{d Q_1^2}{Q_1^2} \frac{d Q_2^2}{Q_2^2} 
\frac{d y_1}{y_1} \frac{d y_2}{y_2} \, t_1 t_2~
 I_3 \left(2 \sqrt{\frac{\alpha_s N_c}{\pi} Y \ln \frac{Q_1^2}{Q_2^2}} \right) \left(\frac{\ln \frac{Q_1^2}{Q_2^2}}{\frac{\alpha_s N_c}{\pi} Y} \right)^{3/2}
 \nonumber \\
\eqa
We will use 
this expression in the following to evaluate the DGLAP Double Leading
Log cross-section.

It is also useful to compare these results with the cross-section corresponding 
to the exchange of one pair of gluons (see equation \ref{ktfactorization2})
\beqa
\label{eeBorn}
&&\hspace{-1.5cm}d\sigma_{e^+e^-}
 (Q_1^2,Q_2^2;y_1,y_2) = \left(\frac{\alpha_{e.m}}{\pi}\right)^2 
\frac{d Q_1^2}{Q_1^2} \frac{d Q_2^2}{Q_2^2} \frac{d y_1}{y_1} \frac{d y_2}{y_2}
\sigma_{\gamma^*\gamma^*}
 (Q_1^2,Q_2^2;Y) \nonumber\\
&& \hspace{-.5cm}=
\frac{128 \alpha_{e.m}^4}{9\pi}\,\frac{d Q_1^2}{Q_1^2} \frac{d Q_2^2}{Q_2^2} 
\frac{d y_1}{y_1} \frac{d y_2}{y_2}
\int \frac{d \gamma}{2 i \pi} 
 \left[(1-y_1) \frac{h_L(\gamma)}{\gamma} +
 \frac{1 + (1-y_1)^2}{2} \frac{h_T(\gamma)}{\gamma}\right] 
\nonumber \\
&&\hspace{-.5cm} \times \, \left[(1-y_2) \frac{h_L(1-\gamma)}
{1-\gamma} +
 \frac{1 + (1-y_2)^2}{2} \frac{h_T(1-\gamma)}{1-\gamma}\right] \, 
\frac{1}{Q_1^2}
\, \left(\frac{Q_1^2}{Q_2^2}\right)^{\gamma} \,.
\nonumber \\
\eqa
Using the expansion of $h_T$ and $h_L$ around $\gamma=0$,
\begin{eqnarray}
\label{devh}
&~& \frac{h_T(\gamma)}{\gamma} = 
\frac{\alpha_s}{3 \pi} \left[ \frac{1}{\gamma^2} + \frac{7}{6 \gamma}
+ \left( \frac{77}{18} - \frac{\pi^2}{6} \right)
+ \left( \frac{131}{27} - \frac{7 \pi^2}{36} \right) \gamma + O(\gamma^2)
\right] \\
&~& \frac{h_L(\gamma)}{\gamma} = 
\frac{\alpha_s}{3 \pi} \left[ \frac{1}{\gamma} - \frac{1}{3}
+ \left( \frac{34}{9} -  \frac{\pi^2}{6} \right) \gamma
+ O(\gamma^2) \right]  
\end{eqnarray}
and the crossing symmetry
\begin{eqnarray}
\frac{h_{T,L}(1-\gamma)}{1-\gamma}=\frac{h_{T,L}(\gamma)}{\gamma}
\end{eqnarray}
together with the Cauchy theorem, one gets
\begin{eqnarray}
\label{2g}
&~& d\sigma_{e^+e^-}
 (Q_1^2,Q_2^2;y_1,y_2) = \frac{d Q_1^2}{Q_1^2} \frac{d Q_2^2}{Q_2^2} 
\frac{d y_1}{y_1} \frac{d y_2}{y_2} 
 \, \frac{64 (\alpha_{e.m}^2 \alpha_s)^2}{243 \pi^3} \frac{1}{Q_1^2} 
\nonumber \\
&~& \left[ t_1 t_2 \ln^3 \frac{Q_1^2}{Q_2^2} + \left( 7 t_1 t_2 
+ 3 t_1 l_2 + 3 t_2 l_1 \right) \ln^2 \frac{Q_1^2}{Q_2^2} \right. \nonumber
\\ &~&  \left.
+ \left( ( \frac{119}{2} - 2 \pi^2 ) t_1 t_2
+ 5 (t_1 l_2 +t_2 l_1) + 6 l_1 l_2 \right) \ln \frac{Q_1^2}{Q_2^2} \right.
\nonumber
\\ &~& \left.
+ \left( \frac{1063}{9} - \frac{14}{3} \pi^2 \right) t_1 t_2
+ (46 - 2 \pi^2) (t_1 l_2 + t_2 l_1) - 4 l_1 l_2 \right] \,.
\end{eqnarray} 

Note that the previous formula was already obtained in reference \cite{bhs}
in the transverse case. The 2-gluon cross-section is an exact calculation
in the high energy approximation
and contains terms up to the NNNLO. The Leading Order part of the 2-gluon
cross-section consists in taking only the $\ln^3 Q_1^2/Q_2^2$ term into 
account.

\section{Numerical Results in the Leading Log Approximation}

In this section, results based on the calculations developed above 
will be given for 
 LEP ($190 \GeV$ centre-of-mass energy) and  a future Linear Collider 
($500 - 1000 \GeV$ centre-of-mass energy).
\gaga interactions are selected at $e^+e^-$ colliders by detecting the 
scattered electrons, which leave the beampipe, in forward
calorimeters. Presently at LEP these detectors can measure electrons 
with an angle $\theta_{tag}$ down 
to approximately 30 mrad. For the LC it has been argued~\cite{brl}
that angles as low as 20 mrad should be reached. Presently~\cite{cdr}
angles down to 40 mrad are foreseen to be instrumented for a generic
detector at the LC. 

Apart from the angle the minimum energy $E_{tag}$ 
for a detectable (tagged) electron is 
important, which is generally dictated by the background conditions
at the experiment. Pair
production background at the LC will make it difficult to measure single
electrons with an energy below 50 GeV. At LEP electrons down to about half of 
the beam energy can be measured.
The energy of the photons $E_{\gamma }$ determine the hadronic
energy of the collision $W_{\gamma\gamma}^2=4E_{\gamma 1}E_{\gamma 2}$,
which should be as large as possible for the test of BFKL dynamics.
In particular the energy dependence of the cross-section is of
interest.
The virtuality $Q_i^2$ of the photon is related to the energy and 
angle of the scattered electron as $Q^2= 4 E_bE_{tag}\sin^2(\theta_{tag}/2)$,
with $E_b$ the beam energy.

 After having specified a region of validity
for our calculations, we will give the accessible integrated 
cross-section as a function of the detector acceptance,
in terms of the energy and angular range of the tagged leptons. 
As a starting point we will assume detection down to $30 \GeV$ and
$33 \mrad$ at LEP, and $50 \GeV$ and $40 \mrad$ at the LC.

\subsection{Kinematical constraints}
Let us first specify the
region of validity for the parameters controlling
the basic assumptions made in the previous chapter. The main constraints
are required by the validity of the perturbative calculations.
The ``perturbative'' constraints are imposed by considering only photon 
virtualities $Q^{2}_{1}$, 
$Q^{2}_{2}$ high enough so that the scale $\mu^{2}$ in $\alpha_S$ 
is greater than 3 $\GeV^{2}$. $\mu^2$ is defined 
using the Brodsky Lepage Mackenzie (BLM) scheme \cite{blm}, 
$\mu^2=\exp(- \frac{5}{3} \sqrt{Q_1^2 Q_2^2})$ \cite{bhs}.
In this case 
$\alpha_S$ remains always 
small enough such that the perturbative calculation is valid. 
In order that gluon contributions dominates the QED one, $Y$ 
(see equation (\ref{Y})) 
is required to stay
larger than $\ln(\kappa)$ with $\kappa = 100.$ (see Ref. \cite{bhs} for 
discussion).
Furthermore, in order to suppress DGLAP evolution, while maintaining
BFKL evolution  will constrain  $0.5 < Q_1^2/ Q_2^2 < 2$
for all nominal calculations.

\subsection{BFKL and DGLAP differential cross-sections}
In this chapter, we will consider the 
DGLAP and BFKL differential cross-sections in 
$y_1$, $y_2$, $Q_1^2$, and $Q_2^2$. It is often assumed \cite{brl} that
the Born cross-section (the exchange of one pair of gluons) is comparable
in magnitude to the DGLAP prediction, since we generally select 
regions where $Q^{2}_{1}/Q^{2}_{2}={\cal{O}}(1)$ in order to observe a large 
BFKL over DGLAP cross-section ratio. In this
domain the DGLAP prediction is expected to be low, as the $k_T$ ordering
required by the DGLAP evolution equation will force the DGLAP cross-section
to vanish if $Q_1 \sim Q_2$.

\begin{figure}
 \begin{center}
   \mbox{\epsfig{figure=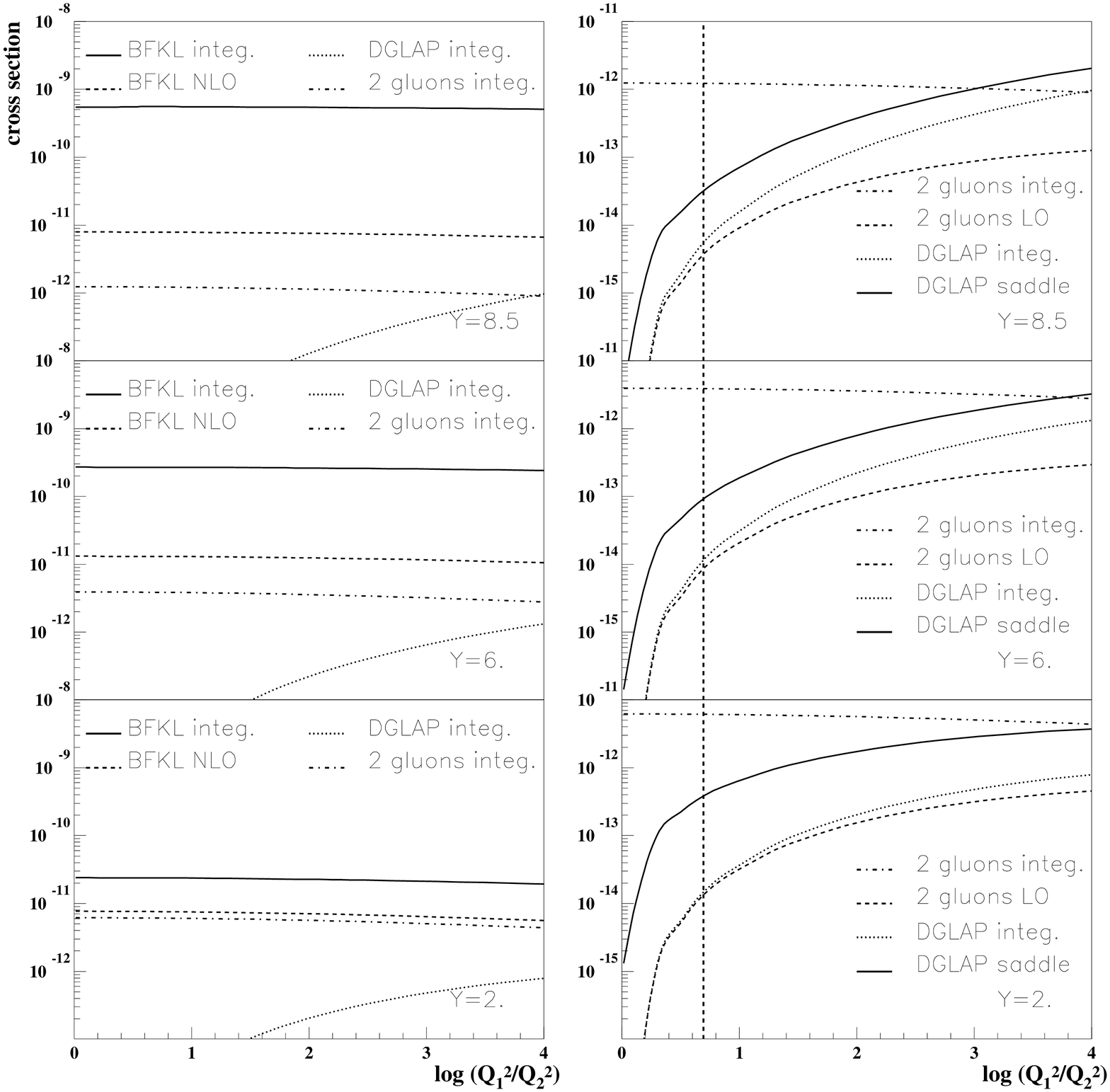,height=160mm}}
 \end{center}
 \caption{Differential cross-sections vs. $\ln Q^{2}_{1}/Q^{2}_{2}$, for 
 different values of $Y$. Exact values are shown
          as well as saddle-point approximations. The dashed vertical
	  line on the left hand side is the value $Q^{2}_{1}/Q^{2}_{2}=2$.}
 \label{diffcs}
\end{figure}

Figure \ref{diffcs} shows the differential cross-sections in the 
BFKL, DGLAP Double Leading Logarithm (DLL) and
2-gluon approximation, 
as a function of $\ln Q^{2}_{1}/Q^{2}_{2}$ and for three values of $Y$. 
The cross-sections on the left hand side 
are calculated using  the unintegrated exact formulae \ref{eeBFKL},
\ref{eeDGLAPb2}, and \ref{2g} for respectively the BFKL, DGLAP (in the double
Leading Log approximation) and 2-gluon exchange cross-sections.
Also the phenomenological NLO BFKL cross-sections, as detailed 
in section 4, are given.

We note that the 2-gluon cross-section is almost always
dominating the DGLAP one in the Double Leading Log approximation.
The saddle point approximation turns out to be  a very good approximation
for the BFKL cross-section and is not displayed in the figure
(saddle-point results are close to the exact calculation
up to $5\%$ in the high $Y$ region, and up to 10\% at lower $Y$. A similar
conclusion was reached in \cite{brl}).
We  note that the difference between the BFKL and 2-gluon 
cross-sections increase with $Y$. 
\par
On the right side of Figure \ref{diffcs}, 
curves for the exact LO and saddle-point 
(eq. \ref{eeDGLAP}) DGLAP calculations are 
shown, as well as the full NNNLO (eq. \ref{2g}) result and the LO (eq. \ref{2g}, 
$\ln^3 Q^2_1/Q^2_2 $ term only)
result for the 2-gluon cross-section. 
Unlike for the BFKL calculation, for the DGLAP
case the saddle-point approximation appears to be in worse agreement
with the exact calculation, and overestimates the cross-section by one 
order of magnitude, which is due to the fact that
we are far away from the asymptotic regime. The comparison between the
DGLAP-DLL and the 2-gluon cross-section in the LO approximation shows that
 both cross-sections are similar when
$Q_1$ and $Q_2$ are not too different (the dashed line describes the
value $Q_1^2/Q_2^2=2$), so
precisely in the kinematical domain where the BFKL cross-section is 
expected to dominate. However,
when $Q_1^2/Q_2^2$ is further away from one, the LO 2-gluon cross-section 
is lower than the DGLAP one, especially at large $Y$. This suggests that
the 2-gluon cross-section could be a good approximation of the
DGLAP one if both are calculated at NNNLO and restricted
to the region where 
$Q_1^2/Q_2^2$ is close to one. In this paper we will use the
exact NNNLO 2-gluon cross-section in the following to evaluate the 
effect of the  non-BFKL background,
 since the 2-gluon term appears to constitute 
 the dominant part of the DGLAP cross-section in the 
region $0.5 < Q_1^2/Q_2^2 < 2$.

\subsection{Integrated cross-sections}
In this chapter, we will study the integrated cross-section over the four
kinematical variables $y_1$, $y_2$, $Q_1^2$, and $Q_2^2$, 
for the exact 2-gluon calculation 
and the saddle point
approximation for the BFKL one. 

First we study the effect of the choice of parameters to define the 
perturbative region for our calculations:
Table \ref{t0.1} shows the effect of varying the cut
on $\mu^{2}$. At the LC, no effect is seen: scattering the incoming leptons
above 40 \mrad requires high photon virtualities so that the 
selected region is  always in the perturbative domain.
Table \ref{t0.2}
contains the  BFKL and 2-gluon 
cross-sections for different values of $\kappa$. 
The BFKL to 2-gluon ratio is  enhanced at
high $\kappa$.
Table \ref{t1.1}
contains the  BFKL and 2-gluon 
cross-sections for different values of the range in ratio $Q_1^2/ Q_2^2$. 
The BFKL to 2-gluon ratio  is rather insensitive to this
restriction. The cut on $1/2 \leq Q_1^2/ Q_2^2 \leq 2$ 
also guarantees that the DGLAP contribution can be well
approximated by the two gluon contribution.

We note that for the parameter choice in this paper
the maximum ratio between the 2-gluon
and BFKL cross-sections is about 20 for the nominal energy and angle cuts 
at the LC and 40 at LEP. 

\begin{figure}
 \begin{center}
  \begin{tabular} {cc}
   \epsfig{figure=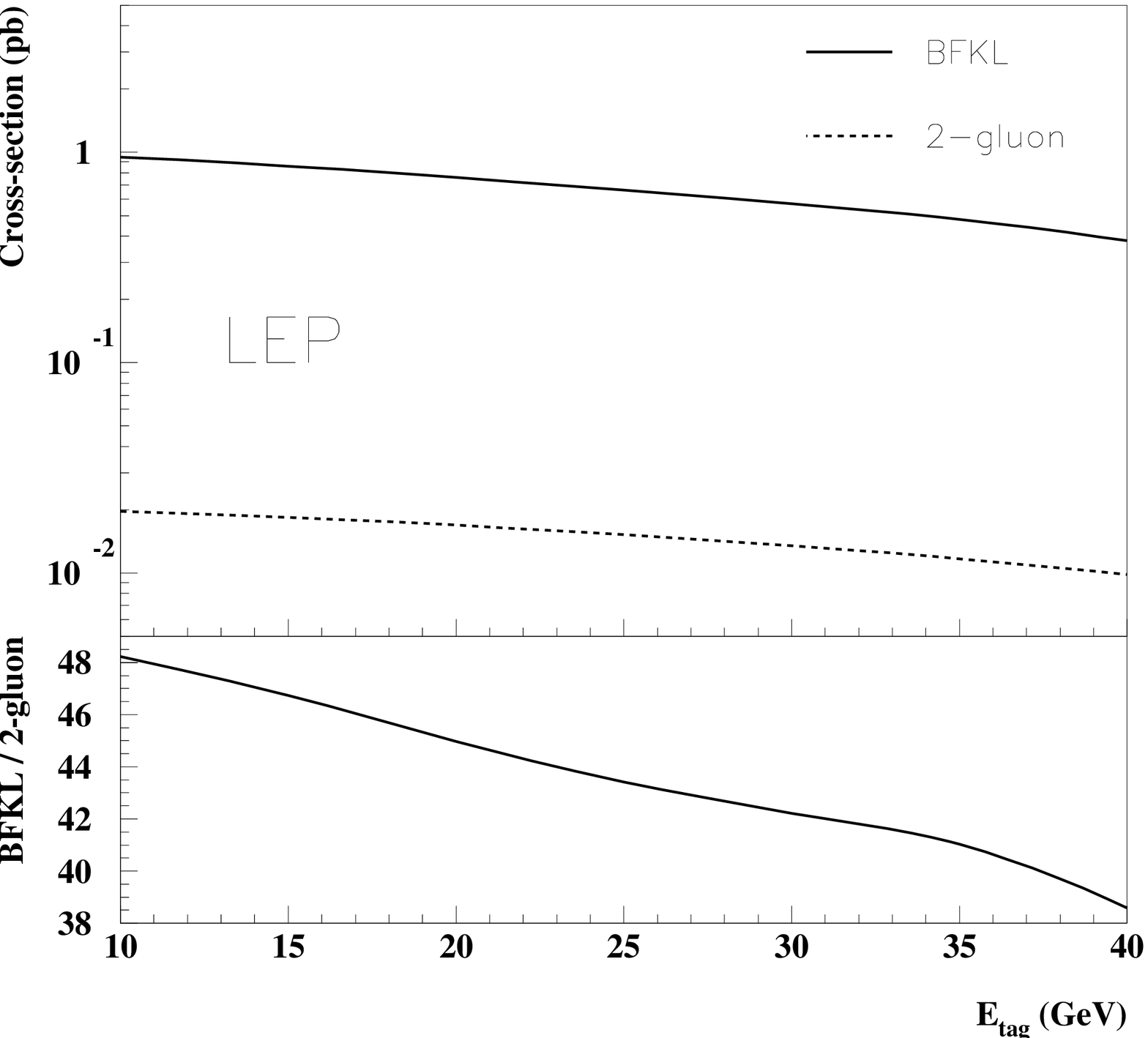,width=80mm} &
   \epsfig{figure=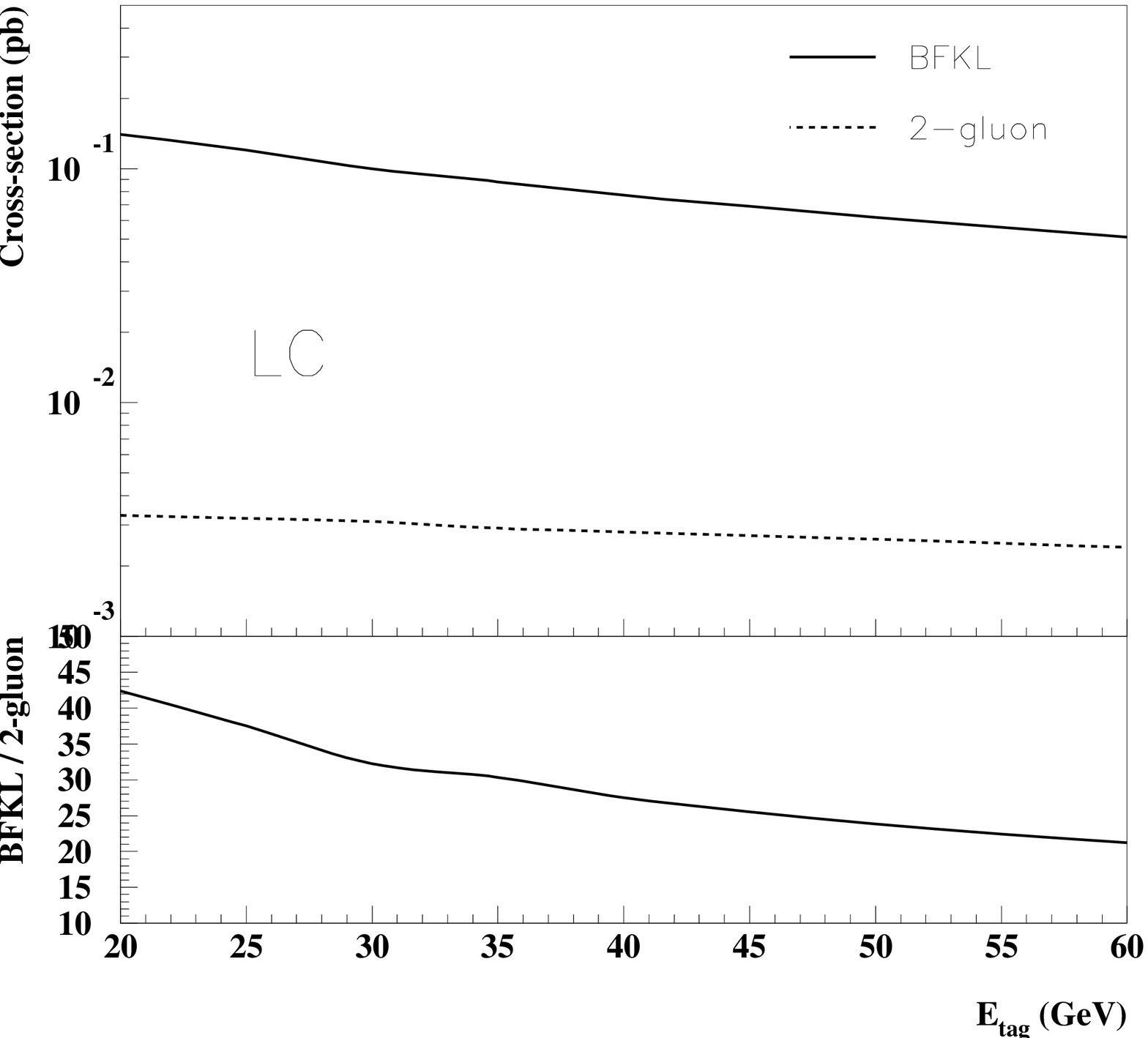,width=80mm}
  \end{tabular}
 \end{center}
 \caption{Integrated BFKL and 2-gluon cross-sections, at LEP and the LC. Leptons are tagged from $E_{tag}$ up to the beam energy.
          We take $\theta_{tag}>33\mrad$ at LEP, $\theta_{tag}>40\mrad$ at the LC.}
 \label{etag}
\end{figure}

\begin{figure}
 \begin{center}
   \mbox{\epsfig{figure=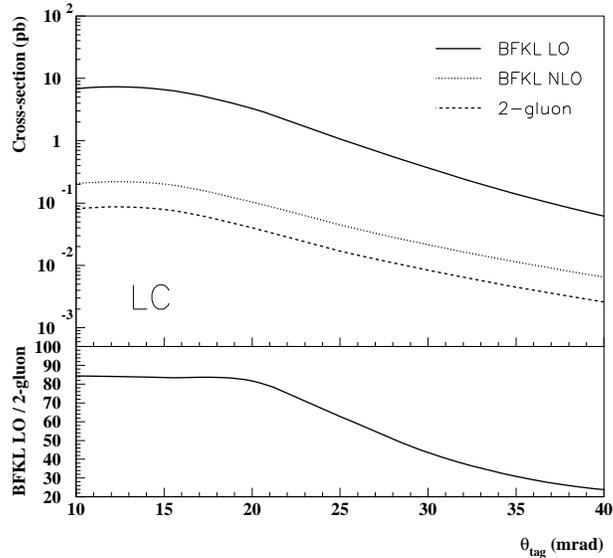,width=80mm}}
 \end{center}
 \caption{Integrated BFKL and 2-gluon
 cross-sections at the LC, for various acceptances. 
Leptons are tagged between 50 and 250 GeV.}
 \label{theta_lc}
\end{figure}

Next we study the effect of the tagged electron energy and angle.
Figure \ref{etag} shows the importance 
of tagging electrons down to low energies (see also Table \ref{t1.2b}). 
Reaching $10 \%$ 
of the beam energy or less allows to enhance the counting rates significantly; 
the difference between the BFKL and 2-gluon predictions 
also increases, improving the detectability of BFKL dynamics. 
The effect of increasing the LC detector acceptance 
for electrons scattered under small angles is 
illustrated in Figure \ref{theta_lc} and in Table \ref{t1.2c}. 
The plateau seen at low
angles results from  the kinematical constraints (see paragraph 3.1).
\begin{table}
\begin{center}
\begin{tabular}{|c||c|c|c||c|c|c|} \hline
 $\mu^{2}$ & $\sigma_{BFKL}^{LEP}$ & $\sigma_{2g}^{LEP}$ & Ratio & $\sigma_{BFKL}^{LC}$ & $\sigma_{2g}^{LC}$ & Ratio \\ \hline\hline
         2 & 2.89 & 3.78E-2 & 76.5 & 6.2E-2 & 2.64E-3 & 23.5  \\
         3 & 0.57 & 1.35E-2 & 42.2 & 6.2E-2 & 2.64E-3 & 23.5  \\
         4 & 0.18 & 6.14E-3 & 29.3 & 6.2E-2 & 2.64E-3 & 23.5  \\ \hline
\end{tabular}
\end{center}
\caption{variation of the cut on $\mu^{2} (\GeV^{2})$, for LEP and the LC. The detector acceptance is taken into account.}
\label{t0.1}
\end{table}

\begin{table}
\begin{center}
\begin{tabular}{|c||c|c|c||c|c|c|} \hline
 $\kappa$ & $\sigma_{BFKL}^{LEP}$ & $\sigma_{2g}^{LEP}$ & Ratio & $\sigma_{BFKL}^{LC}$ & $\sigma_{2g}^{LC}$ & Ratio \\ \hline\hline
        10 & 1.23 & 9.02E-2 & 13.6 & 8.8E-2 & 9.63E-3 &  9.1  \\
        50 & 0.81 & 2.81E-2 & 28.8 & 7.2E-2 & 4.17E-3 & 17.3  \\
       100 & 0.57 & 1.35E-2 & 42.2 & 6.2E-2 & 2.64E-3 & 23.5  \\ \hline
\end{tabular}
\end{center}
\caption{variation of the cut on $\kappa$, for LEP and the LC. The detector acceptance is taken into account.}
\label{t0.2}
\end{table}

\begin{table}
 \begin{center}
  \begin{tabular}{|c||c|c|c||c|c|c|} \hline
  $Q^{2}_{1}/Q^{2}_{2}$&$\sigma_{BFKL}^{LEP}$&$\sigma_{2g}^{LEP}$&ratio&$\sigma_{BFKL}^{LC}$&$\sigma_{2g}^{LC}$&ratio\\ 
  \hline \hline
  0.5-2                 & 0.57            & 1.35E-2        & 42.2 & 6.2E-2          & 2.64E-3           & 23.5  \\
  0.1-10                & 1.71            & 3.94E-2        & 43.4 & 0.123           & 5.65E-3           & 21.8  \\
  0.01-100              & 2.00            & 4.59E-2        & 43.6 & 0.128           & 6.03E-3           & 21.2  \\ \hline
  \end{tabular}
 \end{center}
\caption{Integrated cross-sections (pb) for different ranges of $Q^{2}_{1}/Q^{2}_{2}$, at LEP and LC energies.
         Electrons are detected between 30 and 95 $\GeV$, down to 33 mrad at LEP, and between 50 and 250 GeV, 
         down to 40 mrad at the LC.}
\label{t1.1}
\end{table}

\begin{table}
\begin{center}
\begin{tabular}{|c||c|c|c|} \hline
 $\theta$ & BFKL     & 2-gluon   & ratio \\ \hline \hline
 10       & 6.7     & 8.1E-2   & 82.7  \\
 15       & 6.6     & 7.9E-2   & 83.5  \\
 20       & 3.3     & 4.0E-2   & 82.5  \\
 25       & 1.1     & 1.7E-2   & 64.7  \\
 30       & 0.37    & 8.4E-3   & 44.0  \\
 35       & 0.14    & 4.5E-3   & 31.1  \\
 40       & 6.18E-2 & 2.6E-3   & 23.8  \\ \hline
\end{tabular}
\end{center}
\caption{Integrated BFKL and 2-gluon cross-sections at the LC for different lower
cuts on
$\theta_{tag}$, for the kinematic range defined in the text}
\label{t1.2b}
\end{table}

\begin{table}
\begin{center}
\begin{tabular}{|c||c|c|c|} \hline
 $E$     & BFKL   & 2-gluon  & ratio \\ \hline \hline
 60-250  & 5.1E-2 & 2.4E-3   & 21.2   \\
 50-250  & 6.2E-2 & 2.6E-3   & 23.8   \\
 45-250  & 6.9E-2 & 2.7E-3   & 25.6   \\
 40-250  & 7.7E-2 & 2.8E-3   & 27.5   \\
 35-250  & 8.8E-2 & 2.9E-3   & 30.3   \\
 30-250  & 0.10   & 3.1E-3   & 32.3   \\
 25-250  & 0.12   & 3.2E-3   & 37.5   \\
 20-250  & 0.14   & 3.3E-3   & 42.4   \\ \hline
\end{tabular}
\end{center}
\caption{Integrated BFKL and 2-gluon
 cross-sections at the LC for different lower
cuts on
$E_{tag}$ for the kinematic range defined in the text}
\label{t1.2c}
\end{table}

Finally, in Table \ref{t1.2} we give the measurable cross-sections for 
different values of the beam energy. Although the total
cross-sections increase with $E_{beam}$, the opposite is observed after taking into account the detector acceptance and the fact
that at constant photon virtuality, the scattered electron aligns more and more
 with the beam direction when $E_{beam}$ increases.

\begin{table}
\begin{center}
\begin{tabular}{|c||c|c|c|} \hline
 $E_{beam}$ & $\sigma_{BFKL}$     & $\sigma_{2-gluon}$    & ratio  \\ \hline \hline
 250                & 6.18E-2  & 2.64E-3  & 23.4    \\
 500                & 7.00E-3  & 5.21E-4  & 13.4    \\
 1000               & 8.77E-4  & 9.92E-5  &  8.8    \\ \hline
\end{tabular}
\end{center}
\caption{Integrated cross-sections (pb) for different values of $E_{beam}$ (GeV), after imposing
         $\theta_{tag}>40\mrad$ and $E_{tag}>50\GeV$.}
\label{t1.2}
\end{table}

Assuming an integrated luminosity of $200 \pico^{-1}$ per experiment at LEP with center-of-mass energies around $190 \GeV$, BFKL
predicts roughly 200 events provided forward leptons are tagged down to $10 \GeV$, the 2-gluon prediction being 42 times lower. 
At the LC, with $50 \femto^{-1}$ at $\sqrt{s}=500\GeV$, $E_{tag}>20 \GeV$ and $\theta_{tag}>40 \mrad$, we can expect 7500 BFKL events,
compared to 24 times less for the 2-gluon
contribution. The final results for the cross-sections
are also given in Table \ref{1.fin}. 

\section{Phenomenological approach of NLO effects in BFKL equation}

In this section we adopt a phenomenological approach to estimate the effects
of higher orders. We will generically label these has ``NLO-BFKL'' 
calculations.

\subsection{Variation of the scale for rapidity}
At Leading Order, the rapidity $Y$ is not 
uniquely defined. In the formula
(2.5), it is possible to add a multiplicative constant $\xi$ in front of 
$\hat{s}$. Only a NLO calculation can fix this constant. Taking the
following definition of the rapidity: 
\begin{eqnarray} 
Y = \ln \frac{\xi \hat{s}}{\sqrt {Q^2_1 Q^2_2}}
\end{eqnarray}
we can study the variation of the BFKL and DGLAP cross-sections for different
values of $\xi$. The parameter $\xi$ \cite{bialasb} sets the time
scale for the formation of the interacting dipoles. It defines the
effective total rapidity interval which is $\ln \frac{1}{\sqrt{X_1 X_2}}
+ \ln \xi$, $\xi$ being not predictable (but of order one) at the Leading
Log approximation.
The results are given in Table \ref{t1.10}. We note a large
dependence of the cross-sections on this parameter, and also of the ratio
between the BFKL and 2-gluon predictions which vary between 24 and 2.3!
\par
A phenomenological way to determine this factor $\xi$ has been performed
in Reference \cite{bialas}, where  a four parameter fit of the
proton $F_2$ structure function measured by the H1 collaboration \cite{F2H194}
has been performed using the QCD dipole picture of BFKL dynamics. The parameter
$\xi$ has been found to be 1/3. For this particular value, we note that
the BFKL to 2-gluon ratio prediction is reduced to a value of 12.

\subsection{Effective value for $\alpha_S$}
It has recently been demonstrated 
that the NLO corrections to the BFKL equation are large \cite{lipatov}. 
The main
effect is a reduced value of the so called Lipatov exponent in formula
\ref{eeBFKLa} \cite{gavin}.
A phenomenological way to approach this is to introduce an effective value of
the coupling constant which allows to reduce the value of the Lipatov
exponent. 
\par
In the same 4-parameter fit described above, used to fit inclusive
and diffractive data at HERA, as described in \cite{bialas,nprw},
the value of the Lipatov exponent $\alpha_P$:
\begin{eqnarray}
\alpha_P = 4 \ln 2 \frac{\alpha_S N_C}{\pi}
\end{eqnarray}
was fitted and found to be 0.282. In this fit,  $\alpha_S$
was kept constant. This low value of $\alpha_P$ leads to a low
value of $\alpha_S$ close to 0.11. This low value can be explained
phenomenologically by the decrease of the Lipatov exponent due to large
NLO corrections. 
\par
The same idea can be applied phenomenologically for the $\gamma^* 
\gamma^* $ cross-section. We first studied   the 
variation  of the BFKL
cross-section by setting the scale $\mu^2$ in $\alpha_S$ 
in the exponential of formula \ref{eeBFKLa} to a number and 
consequently taking 
 $\alpha_S$ fixed. The values of $\alpha_S$ and of the BFKL 
cross-sections are given in Table \ref{t1.8}. The cross-section is calculated
for $\mu^2$=10, 100, 1000, and 10000 GeV$^2$ (note that for this study the
term $\sqrt{Q_1^2 Q_2^2}$ is suppressed in the expression of $\mu^2$).
The decrease of the BFKL 
cross-section is quite significant.

\par
The last effect studied  was 
to use a varying $\alpha_S$, and at the same 
time  taking into account the NLO
effects described 
 above. For this purpose, we modify the scale in $\alpha_S$ so that
the effective value of $\alpha_S$ for $Q_1^2=Q_2^2=25$ GeV$^2$ is about
$\alpha_S (M_Z)$. The scale $\mu^2$ for $\alpha_S$ in the exponential is then 
expressed as follows:
\begin{eqnarray}
\mu^2=\zeta  \sqrt{Q_1^2 Q_2^2}.
\end{eqnarray}
The variation of the BFKL cross-section as a function of $\zeta$ are given
in Table \ref{t1.8} for the LC. 

Finally, the results of
the BFKL and 2-gluon cross-sections are given in Table \ref{t1.10} if we
assume both $\zeta$=1000, and $\xi$=1/3 (see paragraph 4.1). 
The $\zeta$ value corresponds to a value of $\alpha_S = 0.11$ for 
$\mu^2 = 10 $ GeV$^2$.
The ratio
BFKL to 2-gluon cross-sections is reduced to 2.3 if both effects are
taken into account together. 

In Table \ref{1.fin}, we also give these effects for LEP with the 
nominal selection and at the 
LC with a detector with increased angular acceptance. 
The ratio given is the comparison of the NLO-BFKL and 2-gluon
cross-section. In both cases the
sensitivity to BFKL effects is increased.
The effect on the cross-section from the angular cut for the LC is shown 
in Figure~\ref{theta_lc}.
The column labelled 'LEP*' gives the results for the kinematic cuts 
used by the L3-collaboration who have recently  presented 
preliminary results\cite{l3}. The cuts are 
$E_{tag}$ = 30 GeV and $\theta_{tag}>30\mrad$ and $\mu^2> 2$ GeV$^2$.
For this selected region the difference between NLO-BFKL and 2-gluon 
cross-section is only a factor of 2.4. A cut on $Q^2_1/Q^2_2$,
as done for the other calculations in this paper,  would help 
to allow a more precise determination of the 2-gluon 'background'.

In Figure~\ref{diffcs} the differential cross-section calculated with the 
BFKL-NLO parameters $\zeta$=10000, and $\xi$=1/3 is shown. In 
Figure~\ref{theta_lc}, the result of the NLO calculation as a function
of $\theta_{tag}$ is also displayed for the same values of $\zeta$ and $\xi$.
An important observation is that the difference between the NLO calculation 
and the LO BFKL calculation in Figure~\ref{diffcs} increases significantly with increasing 
$Y$. Hence to establish the BFKL effects in data, a study of the 
energy or $Y$, rather than the comparison with 
total cross-sections itself, will be crucial. 
Note that an additional uncertainty in the cross-section calculations is the
number of active flavours considered. For this paper 3 quark flavours 
were considered both for BFKL and 2-gluon calculations. Including 
charm only as an additional flavour, without taking into
account any mass effect, would increase both cross-sections by a factor 2.56.
Including charm can be justified for the LC, but some more sophisticated
approach could be necessary for LEP energies.

\begin{table}
 \begin{center}
\begin{tabular}{|c||c|c|c|} \hline
 $\xi$  & BFKL   & 2-gluon  & ratio \\ \hline \hline
 1    & 6.2E-2 & 2.64E-3   & 23.5   \\
 0.1  & 1.6E-2 & 2.64E-3   &  6.1   \\
 0.01 & 6.2E-3 & 2.64E-3   &  2.3   \\
 1/3  & 3.1E-2 & 2.64E-3   & 11.7   \\
 1/3* & 6.2E-3 & 2.64E-3   &  2.3   \\ \hline
\end{tabular}
\caption{Variation of the scale for rapidity (see text). The last number 
(referred by 1/3*) takes also NLO effects on $\alpha_S$ in the
BFKL equation.}
\label{t1.10}
 \end{center}
\end{table}

\begin{table}[tb]
\vspace*{-1.4cm}
\begin{picture}(100,150)(0,0)
\put(105,80){\begin{tabular}{|c||c|c|} \hline
 $\mu^2$      & $\alpha_S$ & BFKL  \\ \hline \hline
 10         & 0.28 & 8.0E-2 \\
 100        & 0.20 & 2.4E-2 \\ 
 1000       & 0.15 & 1.3E-2 \\ 
 10000      & 0.12 & 9.4E-3  \\ \hline
\end{tabular}}
\put(270,80){\begin{tabular}{|c||c|} \hline
 $\zeta$      & BFKL \\ \hline \hline
 $e^{-5/3}$ & 6.2E-2  \\
 10         & 1.3E-2  \\
 100        & 9.4E-3  \\ 
 1000       & 6.2E-3  \\ \hline
\end{tabular}}
\end{picture}
\vspace*{-1.cm}
\caption{Variation of the scale for $\alpha_S$ (the change is made in the 
exponential only). In the left table are given the results for a fixed
$\alpha_S$ (the scale $\mu^2$ is given) 
and in the second table, $\alpha_S$ is running with different values
of the parameter $\xi$ (see text). For
comparison, the 2-gluon cross-section is 2.64 10$^{-3}$ pb.}
\label{t1.8}
\end{table}

\begin{table}
 \begin{center}
\begin{tabular}{|c||c||c|c|c||c|} \hline
   & BFKL$_{LO}$     & BFKL$_{NLO}$ & 2-gluon     & ratio \\ \hline \hline
 LEP        & 0.57   & 3.1E-2       & 1.35E-2  & 2.3     \\
 LEP*        & 3.9   &  0.18       & 6.8E-2  & 2.6     \\
 LC 40 mrad & 6.2E-2 & 6.2E-3       & 2.64E-3  & 2.3     \\
 LC 20 mrad & 3.3    & 0.11         & 3.97E-2  & 2.8     \\ \hline
\end{tabular}
\caption{Final cross-sections (pb), for selections described in the text.}
\label{1.fin}
 \end{center}
\end{table}

\section{Conclusion}
In this paper, we have studied the differences between the 2-gluon and BFKL
and DGLAP
$\gamma^* \gamma^*$ cross-sections 
both at LEP and LC. 
It turns out that the double leading
logarithmic approximation of the DGLAP cross-section 
is much lower than the 2-gluon
one, calculated to NNNLO.
The LO BFKL cross-section is much larger than the 2-gluon cross-section.
Unfortunately, the higher order  corrections 
of the BFKL equation (which we estimated phenomenologically)
are large, and the 2-gluon and BFKL-NLO cross-sections are 
reduced to a factor two to four.
In particular the $Y$ dependence of the cross-section should remain 
sensitive to BFKL effects, even in the presence of large higher 
order corrections.

\section{Acknowledgements}
We like to thank R. Peschanski and J. Bartels and C. Ewerz for many useful
discussions. S.W. thanks the Alexander von Humboldt Foundation
and the II. Institut f\"ur Theoretische Physik, where the first stage of this
work has been done, for support.

\eject

\end{document}